\documentclass[useAMS,usenatbib]{mn2e}

\usepackage{times,graphicx}
\usepackage{amsmath}    
\usepackage{amssymb}    

\newcommand\degrees{^\circ}

\newcommand{\etal}{{et al.~}}

\newcommand\ie{{\it i.e.}}
\newcommand\cf{{\it cf.}}

\newcommand\omp{\Omega_{\rm p}}
\newcommand\omtw{\Omega}
\newcommand\omint{\Omega_{\rm 0}}
\newcommand\kin{{\cal V}}
\newcommand\pin{{\cal X}}
\newcommand\len{a_B}

\newcommand\DLwidth{{$w_{\rm dl}$}}
\newcommand\DLslope{{$s_{\rm dl}$}}
\newcommand\DLxmin{{$x_{\rm min}$}}
\newcommand\DLxmax{{$x_{\rm max}$}}

\def\spose#1{\hbox to 0pt{#1\hss}}
\def\gtsim{\mathrel{\spose{\lower.5ex \hbox{$\mathchar"218$}}
     \raise.4ex\hbox{$\mathchar"13E$}}}
\def\ltsim{\mathrel{\spose{\lower.5ex\hbox{$\mathchar"218$}}
     \raise.4ex\hbox{$\mathchar"13C$}}}

\title{The effect of dust on Tremaine-Weinberg measurements}

\author[Joris Gerssen and Victor P. Debattista]
{Joris Gerssen$^{1}$\thanks{E-mail:jgerssen@aip.de} 
and Victor P. Debattista$^{2}$\footnotemark[1]\thanks{Brooks Prize Fellow}\\
$^{1}$Astrophysical Institute Potsdam, Potsdam D-14482, Germany\\
$^{2}$Astronomy Department, University of Washington, Box 351580, 
Seattle WA 98195}
\begin{document}

\date{Accepted. Received}

\pagerange{\pageref{firstpage}--\pageref{lastpage}} \pubyear{2002}

\maketitle

\label{firstpage}

\begin{abstract}
We investigate the effect of dust on the observed rotation rate of a
stellar bar.  The only direct way to measure this quantity relies on
the Tremaine \& Weinberg method which requires that the tracer
satisfies the continuity equation.  Thus it has been applied largely
to early-type barred galaxies.  We show using numerical simulations of
barred galaxies that dust attenuation factors typically found in these
systems change the observed bar pattern speed by 20-40 percent.  We
also address the effect of star formation on the TW method and find
that it does not change the results significantly.  The results
presented here suggest that applications of the TW method can be
extended to include barred galaxies covering the full range of Hubble
type.
\end{abstract}


\section{Introduction}

The rate at which a bar rotates, its pattern speed, $\omp$, is the
principle parameter controlling a barred (SB) galaxy's dynamics and
morphology.  Most determinations of this parameter are indirect. An
often used method is to match hydrodynamical simulations of SB
galaxies to observed velocity fields \citep[e.g.][for recent
examples]{wei01, per04, rau05}.  Another commonly used method is to
identify morphological features, such as rings, with resonance radii
to infer the pattern speed.

The only direct and model-independent technique to measure $\omp$ is
the Tremaine \& Weinberg (1984, hereafter TW) method.  They show that
if a tracer satisfies the continuity equation then it is
straightforward to derive an expression that relates $\omp$ to the
luminosity-weighted mean velocity, $\kin$, and the luminosity-weighted
mean position, $\pin$, of this tracer.  Gas is generally not a
suitable tracer because it is easily shocked, changes states or is
converted into stars (e.g. Hernandez et al. 2005; see also Rand \&
Wallin 2004).  The assumption underlying the TW method, that the
observed intensity is proportional to the tracer's density, limits its
use to relatively dust and gas-free systems, \ie\ early-type galaxies.
The model-dependent techniques, on the other hand, rely on the
presence of gas and are therefore restricted to late-type barred
galaxies.  There is therefore no overlapping range in Hubble type
where both methods have been applied and compared.

The number of successful applications of the TW method was initially
rather limited: NGC 936 \citep{ken87, mk95}, NGC 4596 \citep{ger99},
NGC~1023 \citep{deb02}.  More recently, larger samples have been
obtained: \citet{agu02} present a study of five SB0 galaxies and
\citet{ger03} study four early-type barred galaxies.  The first
successful application of the method using integral field
(Fabry-Perot) absorption-line spectroscopy was presented by
\citet{dw04} for NGC 7079, while the first direct detection of
kinematic decoupling in a double-barred galaxy, NGC 2950, using the TW
method was presented by \citet{cor03}.  A surprising result emerging
from these studies is that bars rotate as fast as they physically can
\citep[see also][]{rau05}.

Studies of dust extinction in disc galaxies have a long and checkered
history \cite[e.g.][and references therein]{hol05}.  In this paper we
seek to quantify the effect of dust absorption on the TW method.  To
that end we will use the $N$-body simulations of barred galaxies by
\citet{deb03} and \citet{deb06} and implement dust extinction by
adjusting the weights of individual particles.  Our motivation
for this study is to explore how important dust is for TW measurements
of both early and especially late-type galaxies.  In the latter, star
formation can be a potential problem as it violates the basic
assumption of the TW method.  We therefore also explore the ability to
measure $\omp$ in a hydrodynamical simulation of a barred galaxy that
includes star formation \citep{deb06}

Barred galaxies often display prominent dust lanes along the leading
edges of the bar.  Numerical simulations associate the dust lanes with
shocks in the gas streaming along the length of the bar \citep{ath92}.
Observations of velocity jumps across dust lanes, as well as the
detection of dust lanes in radio continuum, confirm this
interpretation \citep[e.g.][]{mun99}.  With the introduction of
efficient spectrometers operating in the IR the possibility now exists
to carry out TW measurements in late-type galaxies, eliminating an
important observational bias.  We therefore test whether such
measurements would be compromised by dust extinction and star
formation.

\begin{figure*}
 \begin{minipage}{140mm}
  \includegraphics[width=7cm]{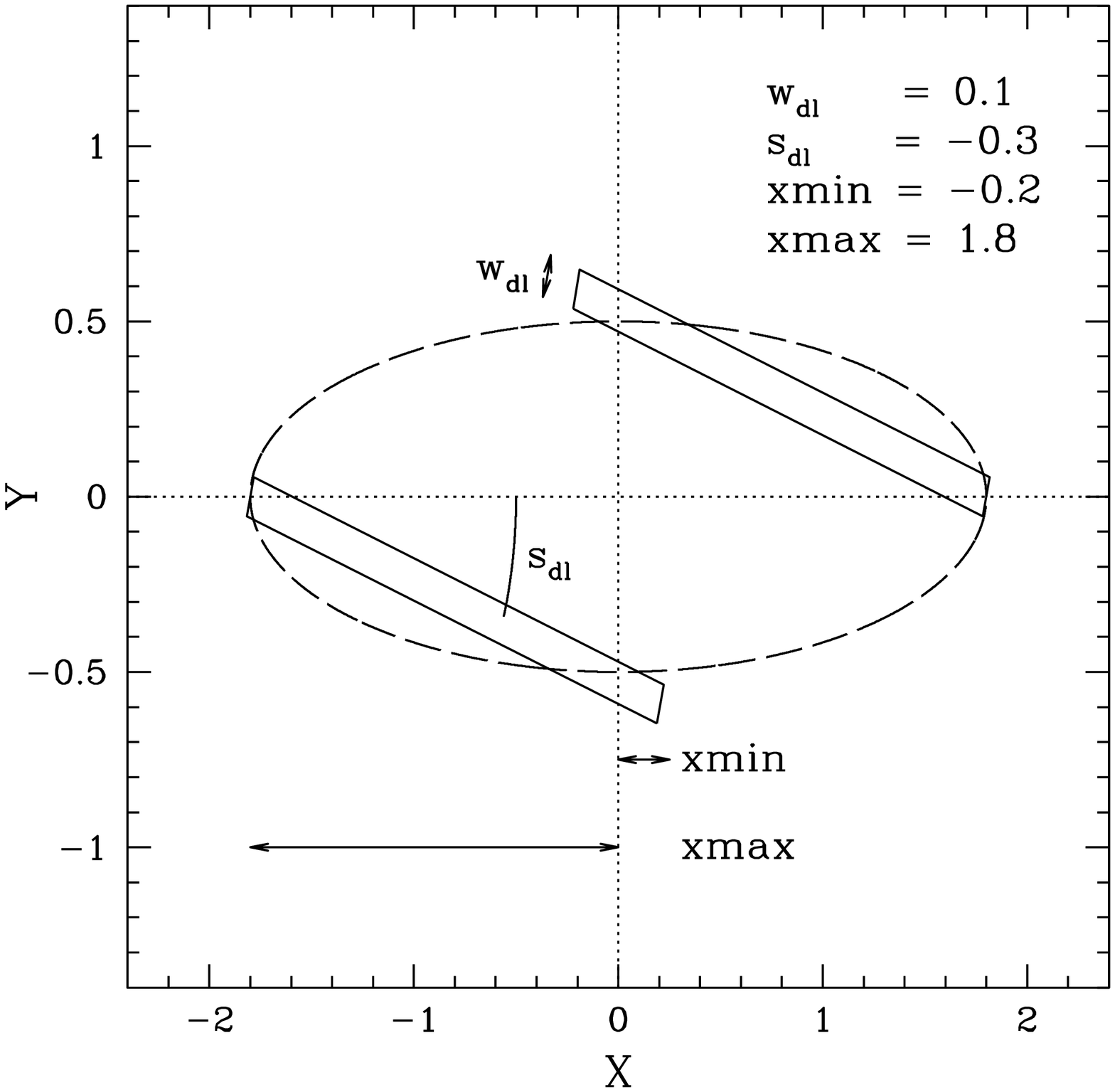}
  \includegraphics[width=7cm]{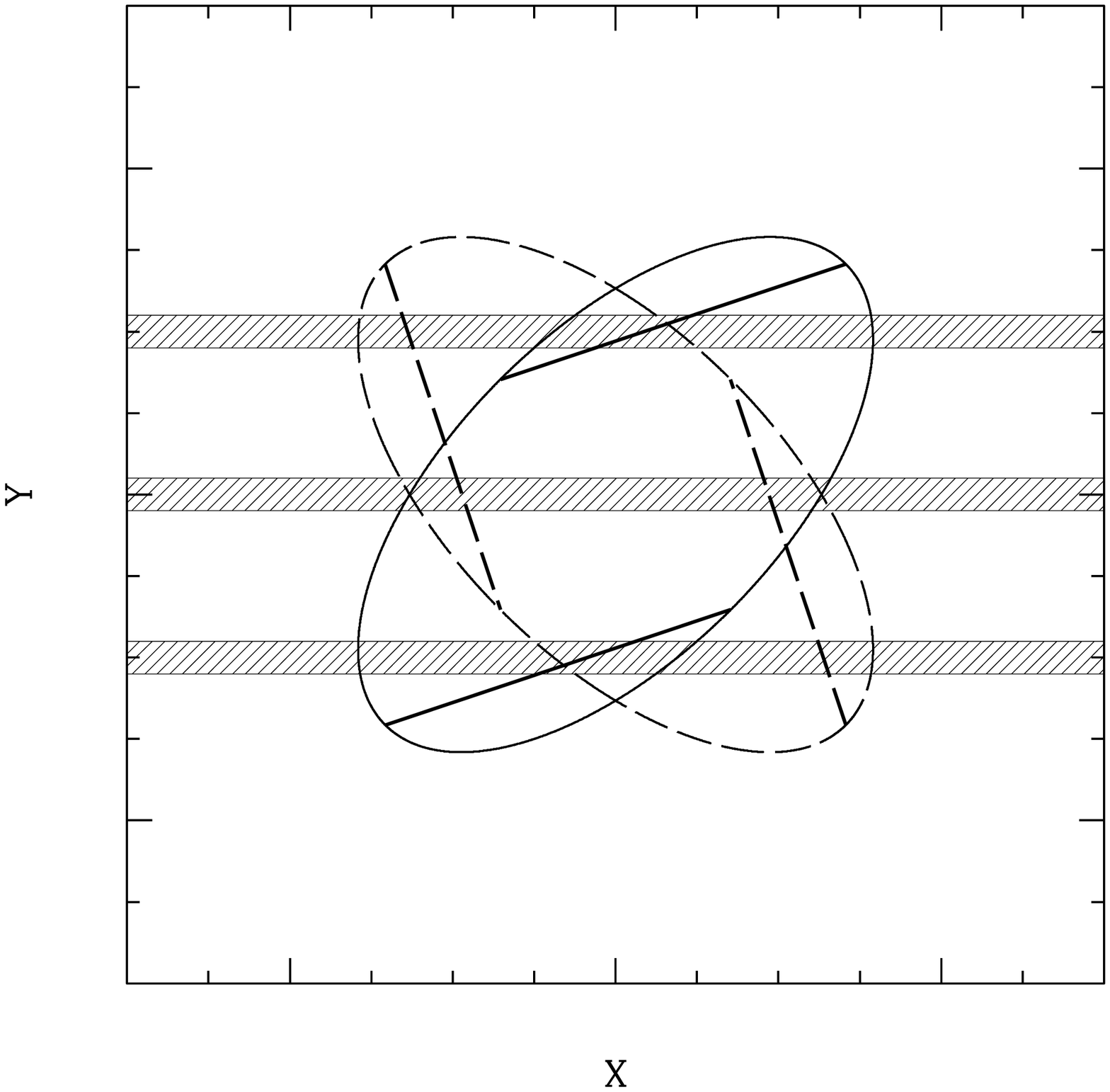}
 \end{minipage}
  \caption{ {\bf left:} panel schematically illustrates the straight
  dust lane geometry (cf Figure~1 in \citet{ath92}) used in our
  numerical experiments. The bar is represented by the dashed ellipse
  in this cartoon.  {\bf right:} This cartoon shows that TW
  measurements of the bar at a positive PA (solid lines) are not
  equivalent to the same observations at a negative PA (dashed lines).
  This is because the long-slits (hatched areas) intersect the dust
  lanes, represented by the straight line segments, at different
  locations depending on the sign of the bar position angle, which
  leads to a change in sign of the error in $\omtw$.
  \label{f:dustlane}}
\end{figure*}

Our results show that dust lanes on the leading edges of a bar change
the TW derived value of $\omp$ depending on the geometry of the
observations.  This can lead to measured pattern speeds that differ by
40 percent or less from their true value for realistic dust lane
models. As this change can be rather small and because the extinction
can be observationally estimated, it suggests that the TW method can
be applied to the full range of Hubble type.

The paper is organized as follows. We summarize the TW method and
describe the models to which we apply it in Sections 2 and 3
respectively.  In Section 4 we describe our models for the dust lane
and the diffuse dust disc. The model predictions are presented in
Section 5 and in Section 6 we compare the models to data and discuss
the implications of our modeling.

\section{TW Method}
\label{s:twmethod}

For any tracer population that satisfies the condition of continuity,
it can be shown that
\begin{equation}
\kin = \pin\omp \sin i.
\label{e:tweqn}
\end{equation}
Here, $\pin \equiv \int h(Y) X \Sigma \ dX dY$ and $\kin \equiv \int
h(Y) V_{\rm los} \Sigma \ dX dY$ are luminosity-weighted positions and
velocities respectively.  The surface brightness of the tracer
population is $\Sigma$, and $V_{\rm los}$ is the line-of-sight
velocity minus the systemic velocity.  The galaxy's inclination is
given by $i$ and coordinates along its major and minor (disc) axes by
$(X,Y)$.  The weighting function, $h(Y)$, can be chosen arbitrarily 
(TW).  For several slit observations (\ie\ $h(Y) \propto
\delta(Y-Y_{\rm offset})$), plotting $\kin$ versus $\pin$ produces a
straight line with slope $\omp \sin i$.  In our experiments, we will
refer to the slopes obtained from fitting such straight lines as
$\omtw \sin i$, in order to distinguish from the intrinsic pattern
speed $\omp$ obtained from the time evolution.

We apply the TW method to the $N$-body models of barred galaxies
described in Section \ref{s:nbody} as done by Debattista (2003,
hereafter D03).  In our measurements, except where noted, we set the
inclination $i = 45\degrees$ and the projected angle between the bar
major axis and disc major axis, PA$_{\rm bar} = 45\degrees$, implying
an intrinsic angle of $54.7\degrees$, which is a favorable orientation
for the TW method.

As in D03 we use 11 slits each with a width of 0.1 (initial) disk
scale-lengths \citep[the units of the simulation are described
in][]{deb06} to derive $\omtw$. Tests with slits half and twice as
wide return nearly identical results.  Applications of the TW method
to real data generally use between three and five long-slit spectra.
(Integral field observations could soon change this.)  A typical
observational setup places one slit on the major axis and two slits on
opposite sides either near the bar edges, or at half the bar radius.
Tests with such configurations in the simulations did not yield values
of $\omtw$ that were significantly different from the results derived
with all 11 slits.  The strongest constraints on the slope come from
slits located about half a bar radius from the galaxy center.  The
leverage on the slope of the fit derived from spectra beyond this
point is decreasing with radius, as is their signal-to-noise ratio
(\cf\ D03).  Progressively removing pairs of the outermost slits from
the set of 11 and repeating the measurements has little effect on
$\omtw$, although it increases the scatter.

\section{Barred galaxy models}
\label{s:nbody}

In their study of the effect dust has on the observed kinematics of
discs, \citet{bae03} used analytical distribution functions for the
intrinsic stellar kinematics.  However, such methods are ill-suited
for modeling barred galaxies.  In this paper we explore the effect of
dust on barred systems using $N$-body and $N$-body+SPH models.

\subsection{$N$-body models}

Our four $N$-body simulations TW1-TW4 were evolved on a polar
cylindrical grid code \citep{sv97}.  The radial spacing of grid cells
in this code increases logarithmically from the center and reaches to
well past the edge of the model discs.  We used Fourier terms up to
$m=8$ in the potential, which was softened with the standard Plummer
kernel.  Time integration was performed with a leapfrog integrator
using a fixed time-step.

These models use rigid halos; this allowed us to sample the disc (and
bulge, if present) component at high mass resolution.  A full
description of these models can be found in D03 and \citet{deb06}.
Briefly, the rigid halos were represented either by a spherical
logarithmic potential or by a Hernquist halo.  In all models, disc
kinematics were set up using the epicyclic approximation to give a
constant Toomre $Q$.  We assume that all the late-type models are
bulgeless and only include a bulge component (using the method of
Prendergast \& Tomer, 1970) in the early-type model presented in D03.
We use units in which the initial radial scale length of the disc
particle distribution is $1$; in models TW1-TW4 the vertical scale
height is then $0.1$ in these units.  Except where noted, we report
results using these natural units.  The resulting axisymmetric systems
all formed rapidly rotating bars.

The model TW5 is an unpublished simulation with an exponential disk +
dominant live dark matter halo evolved with {\sc pkdgrav}.  Because
the halo was live and dominant in this simulation the bar ends well
inside its co-rotation radius, i.e. it is a slow bar (Debattista \&
Sellwood 1998, 2000).  This also causes the bar to be larger and
stronger than any of the others.

\subsection{$N$-body+SPH}
\label{s:sf}

In late type galaxies the rate of star formation (hereafter SF) can be
substantial.  Thus stars violate the required condition of
(source-free) continuity.  To quantify the influence of SF we also
include an $N$-body+SPH model of a barred galaxy in our analysis.
This model, described in detail in \citet{deb06}, included a live halo
and was evolved using a tree code ({\sc gasoline}).  The star
formation recipe is the one of \citet{kat92}. In our model the SF is
most pronounced in the central regions of the system and in the bar
itself, consistent with numerous observational results, (e.g. Sheth et
al. 2002).

We measured $\omp$ directly from the time-evolution; since the dark
matter halos in these models are rigid, $\omp$ remained constant after
an early phase of evolution.  Table \ref{t:snapshots} lists the
properties of the bars, including the pattern speeds $\omp$, of all
the models.

\subsection{Individual models}

\begin{table*}
\caption{Our numerical models have bar semi-major axis $\len$ The
values of $\len$ were estimated by eye.  We measured the pattern
speed, $\omp$, from the time evolution. To obtain the dust-free TW
method values, $\omint$, we orientated the models to $i = 45 \degrees$
and projected PA$_{\rm bar} = 45 \degrees$. The difference between
$\omp$ and $\omint$ is listed under column Error.
\label{t:snapshots} }
 \begin{minipage}{140mm}
 \centering
  \begin{tabular}{lccccl}
  \hline
   Run & $\len$ & $\omp$ & $\omint$ & Error & Reference\\
       & kpc    & km s$^{-1}$ kpc$^{-1}$ &   km s$^{-1}$ kpc$^{-1}$ & \\
 \hline
%
TW1 & 4.5 & $23.3 \pm 0.8$ & $26.5 \pm 0.8$ & $+14\%$ & \citet{deb03}\\
TW2 & 7.8 & $15.0 \pm 0.2$ & $15.9 \pm 0.1$ & $ +6\%$ & \citet{deb06}\\
TW3 & 3.8 & $22.8 \pm 0.2$ & $24.2 \pm 0.2$ & $ +6\%$ & \citet{deb06}\\
TW4 & 3.3 & $34.6 \pm 1.1$ & $38.0 \pm 4.7$ & $+10\%$ & \citet{deb06}\\
TW5 & 8.0 & $ 8.8 \pm 0.4$ & $9.5 \pm 0.5$  & $ +8\%$ &  unpublished \\
TW6 & 5.9 & $26.4 \pm 0.2$ & $25.3 \pm 1.1$ & $ -4\%$ & \citet{deb06}\\
\hline
\end{tabular}
\end{minipage}
\end{table*}

The models were selected to have a variety of morphologies and pattern
speeds.  We generally selected late times in the simulations to ensure
that the models are well settled.  Here we use model units.


{\it TW1}. This model of an early-type galaxy was described in D03,
who used it to study the effect of position angle errors on TW
measurements.  We use the model at $t=2.48$ Gyr, by which time a
strong and stable bar had formed. Only weak spirals are present in
this model, resulting in a minimal interference with the determination
of $\omp$.  This model has $4\times 10^6$ equal-mass particles in
bulge$+$disc.

{\it TW2}. The evolution of this model is discussed in \citet{deb06},
where it is referred to as model L5.  We use model TW2 at $t=7.44$ Gyr
at which time its spiral is relatively weak.  It is a late-type model
with $7.5 \times 10^6$ particles.

{\it TW3}.  Model TW3 is described in \citet{deb06}, where it is
referred to as model H2, and in \citet{deb05}, where it is referred to
as R7.  We use $t=2.48$ Gyr at which time spirals are rather weak.
This late-type model has $4\times 10^6$ particles.

{\it TW4}.  This late-type model is described in \citet{deb06} where
it is referred to as model S1.  We use model TW5 at an early time,
$t=1.24$ Gyr, when the spiral arms were prominent.  This model contains
$7.5\times 10^6$ particles.

{\it TW5}. Slow bar model, with $0.75 \times 10^6$ particles in the
disk and $2\times 10^6$ in the halo.  We use this model at $t \sim
10.8$ Gyr.  Unlike the other models, TW5 is not yet published.

{\it TW6}.  This model includes star formation. It is described in
\citet{deb06} where it is referred to as NG3.  We use this model at
$t=3.48$ Gyr.  At this time the model contains $0.7 \times 10^6$
particles.

\begin{figure*}
\includegraphics[width=15cm]{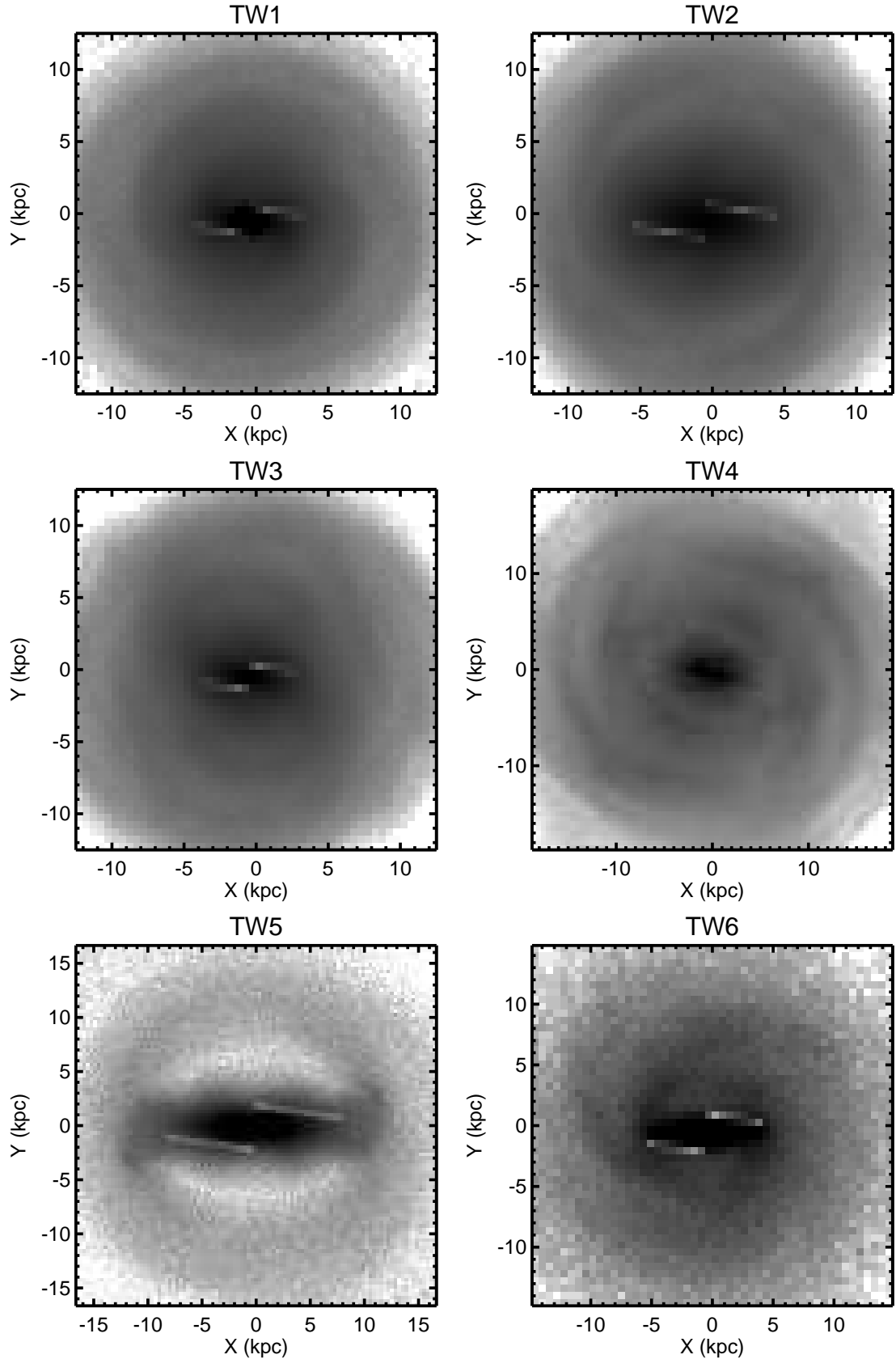}
\caption{Examples of the projected distribution of particle weights in
our numerical models shown on a logarithmic stretch.  The models are
shown here face-on to highlight their morphological features.  Dust
lanes have been included to illustrate how dust attentuation is
implemented in our simulations.  For clarity the dust lanes have an
exaggerated extinction ($A_v = 15$) and width.
\label{f:dustmap}}
\end{figure*}

\section{Dust Models}
\label{s:dustdist}

Observations of dust in disc galaxies show that the dust distribution
in these systems follows a double exponential profile \citep{wai89}
and it is conventional to model them as such \citep[e.g.][]{bae03,
vg04}.  Thus we model diffuse dust discs as:
\begin{equation}
 D_{\rm disc}(R,z) = D_{0,{\rm disc}} \ e^{-R/h_{R,{\rm disc}}} \ 
  e^{-|z|/h_{z,{\rm disc}}}
\label{eq:dustdisc}
\end{equation}
where $h_{R,{\rm disc}}$ and $h_{z,{\rm disc}}$ are the radial
scale-length and the vertical scale-height of the dust disc
respectively.  The central dust density is given by $D_{0,{\rm
disc}}$.

\subsection{Dust Lanes}
\label{s:dustgeom}

Late-type barred galaxies often exhibit prominent dust lanes running
from the end of the bar toward the center of the galaxy.
\citet{ath92} notes that these dust lanes generally fall in two
categories: (i) straight dust lanes which are parallel to each other
but make an angle with the bar major axis, and (ii) curved dust lanes
which are parallel to the bar major axis but offset toward the leading
edges and with their concave side facing the bar major axis.  We
implemented both morphologies in our experiments.  Since the results
obtained with the curved dust lane morphology and the straight dust
lane morphology are qualitatively and quantitatively similar we
concentrate only on the latter in this paper.

We implement dust lanes by defining rectangular areas on the bar's
leading edges that mimic Athanassoula's straight dust lanes.  The dust
lane geometry is fully specified by four parameters: the width of the
dust lane (\DLwidth), the angle the dust lane makes with the bar major
axis (\DLslope), and the minimum and maximum extent of the dust lane
(\DLxmin\ and \DLxmax\ respectively) along the $x$ (bar) axis. In
practice we fixed \DLxmax\ to the radius of the bar and the dust lanes
start on the bar's major axis.  A cartoon representation of these
various parameters is shown in the left panel of
Figure~\ref{f:dustlane}. Note that \DLxmin\ can be negative, \ie\ the
dust lanes can extend past the center of the bar.  In this case, the
inner parts of the dust lanes are sometimes observed to wind around
the bulge, see \citet{ath92}.  The sign of PA$_{\rm bar}$ is
important, as illustrated in the right panel of
Figure~\ref{f:dustlane}, because the slits used to measure the TW
integrals intersect the dust lanes at different locations for positive
and negative PA$_{\rm bar}$.

We assume that the dust distribution within the dust lanes can also be
described by a double exponential model:
\begin{equation}
  D_{\rm lane}(R,z) =
\begin{cases}
D_{0,{\rm lane}} \ e^{-R/h_{R,{\rm lane}}} \ e^{-|z|/h_{z,{\rm lane}}} 
  & \text{if in dust lane,}  \\
0 & \text{otherwise.} \\
\end{cases}
\label{eq:dustlane}
\end{equation}
Although the dust lane model has a similar double exponential profile
as the dust disc model, the scale parameters of the two components
will in general be different.

\subsection{Particle weights}
\label{ssec:weights}

In a particle implementation of the TW method, the integrals in
Eqn. \ref{e:tweqn} are replaced by sums over particles.  Each particle
has its own weight, $w$.  In the collisionless $N$-body models (TW1 to
TW5) all particles are coeval and have equal mass.  We therefore
assign the same intrinsic (i.e. before applying dust attenuation)
weights to all particles in these models

In model TW6, however, the particle luminosities, and hence their,
weights will depend on their age. To account for this we scale the
intrinsic weights of all particles by their M/L ratio in this model.
We do not a priori know the luminosity of a particle. We only know its
mass $M_p$ and time of creation $t_p$.  To calculate the luminosity of
a particle we follow the procedure first described by Tinsley (1973,
see also van~der~Kruit 1989) and summarized in the appendix.

When dust is present, particles obscured by dust contribute to the
integrals with lower weights.
Unlike a foreground screen model, the dust extinction in our models
varies with position within the disc.  For each particle we calculate
the amount of intervening dust by integrating the projected dust
distribution along the line-of-sight.  The weight for each particle is
then given by $w_i = e^{-\tau_i}$, where the optical depth $\tau_i$ is
obtained from
\begin{equation}
 \tau_i = \int^{s_i}_{-\infty} D(s') \ ds',
\end{equation}
assuming a unit mass absorption coefficient.  This integral is
evaluated using a Gauss-Legendre 64-point quadrature formula from the
FORTRAN NAG libraries.  To guard against numerically unstable
solutions we subdivided the integration interval into ten subintervals
to improve the accuracy.  
Examples of the weight distribution in each of our six models are
shown in Fig.~\ref{f:dustmap}. They are shown here face-on to better 
illustrate their morphological features.

Following convention, we specify the amount of extinction in our
models as the face-on optical depth for a particle infinitely far
behind the disc, $\tau_0$.  For the dust disc models, we place this
particle behind the center of the disc, while for the dust lanes we
place it at a point halfway along the dust lane.  Equivalently, the
extinction can be expressed in $A_V$ magnitudes
\begin{equation}
A_V ~=~ 2.5 \log_{10}e~ \tau_0 ~=~ 1.086 \tau_0.
\end{equation}

\subsection{Code test: A dust disc}

\begin{figure}
  \includegraphics[width=8cm]{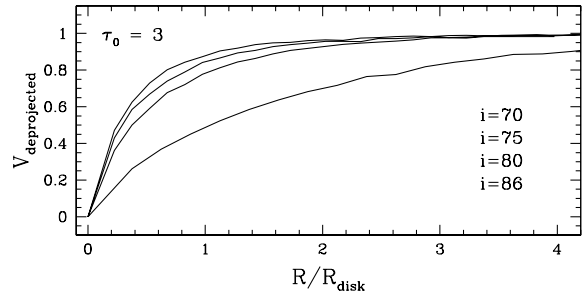}
  \caption{To test our dust code we reproduce the analytical results
   of VG04 and derive the major axis rotation curves in an
   axisymmetric particle data set. We added a diffuse dust disk with a
   face-on central optical depth of $\tau_0 = 3.0$ and derive the
   rotation curve at various inclinations. The projected amount of
   dust extinction varies with inclination leading to more slowly
   rising rotation curves.  From top to bottom the inclination of the
   rotation curves varies between $70\degrees$ and $86\degrees$.  At
   the largest inclination, the rotation curve does not reach its
   asymptotic values before the edge of the $N$-body data (cf
   Figure~1a of VG04).
  \label{f:codetest}}
\end{figure}

\citet[][hereafter VG04]{vg04} presented an analytical model of a
diffuse dust disc and used this model to demonstrate that the presence
of dust lowers the slope of the observed inner rotation curve at large
inclinations.  As a test of our dust extinction code we reproduce
their result using a simple axisymmetric double-exponential model of a
cold rotating disc with $R_{\rm d} = 1.0$ and $z_{\rm d} = 0.1$
respectively.  The circular velocities are implemented using
equation~2 of VG04 and their `luminosity class 1' parameters.  We
sample this model randomly by $10^6$ particles.  Dust in this model is
described by a diffuse disc with geometric parameters that are
identical to those for the particle distribution: $h_{R,{\rm disc}} =
R_{\rm d}$ and $h_{z,{\rm disc}} = z_{\rm d}$.  In order to facilitate
a direct comparison with the results of VG04 we use the same dust
extinction in the disc ($\tau_0 = 3.0$) as they do in their Figure~1a.

We measured the rotation curves in our model by radially binning the
particle velocities in a slit along the major axis.  In each radial
bin the line-of-sight velocity distribution (LOSVD) of the particles
has a sharp cut-off at high velocities and a tail toward lower
velocities.  We take the cut-off velocity as the measure of the
circular velocity in each bin.  The resulting rotation curves are
shown in Fig.~\ref{f:codetest}. As these curves are derived from a
finite number of particles they are not as smooth as the analytical
results of VG04 but their qualitative and quantitative behaviour is
quite similar: internal extinction decreases the slope of the inner
rotation curves.  This agreement with the results of VG04 strengthens
our confidence in our extinction code.

\section{Results}

\begin{figure*}
\includegraphics[width=17cm]{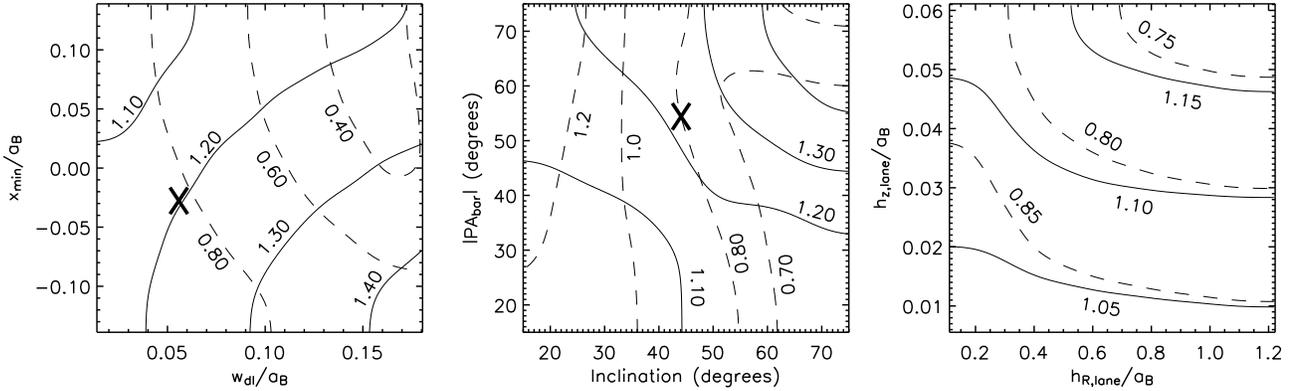}
\caption{Illustrations of the effect of varying the {\it dust lane}
model on $\omtw$.  The contours show the ratio $\omtw / \omint$.  In
each panel we evaluate this ratio at a fixed value of the extinction
($A_V \sim 8)$ and with a dust lane slope of \DLslope\ $= -0.3$.
Unlike the axisymmetric {\it dust disc} model, the sign of PA$_{\rm
bar}$ matters (see Figure~\ref{f:dustlane}). The plotted contours are
for PA$_{\rm bar} = +45\degrees$ (solid) and PA$_{\rm bar} =
-45\degrees$ (dashed).
{\bf left:} The ratio $\omtw / \omint$ for varying dust lane width
(\DLwidth) and minimum dust lane radius (\DLxmin).
{\bf centre:} The behaviour of the ratio $\omtw / \omint$ as a function
of viewing parameters.  In this panel the dust lane model parameters
are fixed at \DLxmin = $-0.1$ and \DLwidth = $0.1$.  
{\bf right:} The dependence of the ratio on the scale length
parameters of the dust lane model. For reference, the initial scale
lengths of the disc particle distribution are $R_{\rm d} = 1.0$ and
$z_{\rm d} = 0.1.$
The cross in the left and middle panels marks the approximate location
of NGC~4123 in these panel (see Section~\ref{s:discussion} but note
that the extinction in NGC~4123 is only $A_V \simeq 3$).
\label{f:parspace} }
\end{figure*}

The TW-measured value of the bar pattern speed, $\omtw$, can be
affected not just by dust but also by spiral structure.  We quantify
the effect of dust using the ratio $\omtw / \omint$, where $\omint$ is
the TW measured value in the absence of dust.  To quantify the effect
of spirals, we compare $\omint$ with $\omp$, the pattern speed
inferred from the time evolution of the models, in the absence of
dust.  The resulting values of $\omint$ for each $N$-body model are
listed in Table~\ref{t:snapshots} and compared with $\omp$.  The TW
measurements generally differ from the time evolution values by
$\ltsim 15\%$ (as already noted by D03).  Model TW5 included strong
spirals in the disc; despite this, $\omtw$ differs from $\omp$ by less
than $10$ percent.

We conclude that in the presence of a strong bar, spirals only perturb
TW measurements rather weakly.  This is presumably because spirals are
at large radius and therefore in lower density parts of the disc.
Moreover, because spirals wind around the galaxy centre, their
contribution to any slit partly cancels out.  Therefore, spirals are
not expected to present a serious problem for TW measurements in
late-type galaxies with strong bars.  (But note that, other than
obscuration, we give all particles equal weight; in a real spiral star
formation would increase its relative contribution in blue bands.)  In
the remainder of this paper, we will compare the measurements with and
without dust directly, rather than with the time evolution value.

\subsection{Diffuse dust disc}

We first briefly consider the effect on $\omtw$ of a diffuse dust
disc.  We examined the ratio $\omtw / \omint$ resulting from varying
the parameter $D_{0,{\rm disc}}$, and therefore $A_V$.  Typically
observed values for $A_V$ range from 0.5 to 4.0 mag \citep{hol05}.
The change in $\omtw$ is only five percent for $A_V \sim 3$.  Even at
$A_V=8$, the change is less than 15 percent.  We also examined the
dependence on the assumed $h_{R,{\rm disc}}$ and $h_{z,{\rm disc}}$ of
the dust disc at a constant $\tau_0 = 5$. Unsurprisingly, the value of
$\omtw / \omint$ depends only weakly on either scale parameter at
fixed $\tau_0$ once they become of order of, or larger than, the scale
of the stellar system itself.  Even for $\tau_0 = 5$, the error in
$\omtw$ does not exceed 10 percent.

In real galaxies, dust within the bar radius is likely to be swept
into a dust lane.  To simulate this we tested what the effect of a
diffuse dust disc with a hole within the bar radius is
\citep[cf][]{mar06}.  We set $h_{R,{\rm disc}} = 10$, $h_{z,{\rm
disc}} = 0.1$, and $\tau_0 = 10$.  As intuitively obvious, such a dust
disc has a negligible ($< 2$ percent) effect on $\omtw$.

Diffuse dust discs lead to a small errors in $\omtw$, of order $10$
percent.  Existing TW measurement of early-type barred galaxies that
lack dust lanes but possibly have diffuse dust discs are therefore
unlikely to have been affected substantially by dust.  All bars with
TW measurements so far are consistent with being fast, in the sense
that they end near their corotation radius.  This finding supports low
dark matter densities at the centers of galaxies \citep{ds98, sd06}.

\subsection{Dust lanes}
\label{s:dustlane}

We therefore ignore the dust disc contribution in our dust lane models
and start by exploring the dust lane geometry and the dust parameters
using model TW1.  This is a strongly barred model.  As strongly barred
galaxies are observed to have flat surface brightness profiles along
their major axis \citep{elm96}, we set $h_{R,{\rm lane}} = 10$ to
obtain a nearly constant radial dust lane distribution.  For the
vertical scale height of the dust lane we set a value of $h_{z,{\rm
lane}} = 0.1$, which is comparable to the initial stellar
distribution.

Using this distribution for the dust lane and $D_{0,{\rm lane}} = 100$
(corresponding to an extinction of $A_V \sim 8$) we evaluate the ratio
$\omtw / \omint$ for varying \DLslope, \DLwidth\ and \DLxmin.  The
solid lines in the left panel of Figure~\ref{f:parspace} illustrate
the variation in this ratio as a function of \DLwidth\ and \DLxmin\
for constant \DLslope\ $=-0.3$.  For all combinations of parameters
the derived ratio is larger than one.  Dust lanes therefore lead the
TW method to overestimate the pattern speed by 10 to 40 percent.
Variations in the width have the largest impact on $\omtw$ as they
affect the geometric area of a dust lane the most.

As noted in Section~\ref{s:dustgeom} the (observationally known) sign
of PA$_{\rm bar}$ is important for the dust lane model.  We illustrate
this in the same panel using dashed contours for PA$_{\rm bar} =
-45\degrees$.  The effect of a negative PA is that $\omtw$ becomes
smaller than $\omint$ rather than larger as it does for positive PA.
Dust lanes can thus also lead the TW method to underestimate the
pattern speed.  For realistic values of the model parameters (e.g. the
NGC~4123 values marked by the cross, see also
section~\ref{s:discussion}) the percentage change in $\omtw / \omint$
is roughly independent of the sign of PA$_{\rm bar}$.

We repeated the exercise for a range of dust lane slopes bracketed by
$-0.4\leq$ \DLslope\ $\leq-0.1$.  The behavior of $\omtw / \omint$ is
qualitatively and quantitatively similar for all but the most shallow
slope. When \DLslope\ $= -0.1$, the errors in $\omtw$ are rather
small.  This happens because the dust lane is no longer near the
leading edge of the bar but closer to the bar major axis. However
\DLslope = $-0.1$ bears little resemblance to realistic dust lanes.
For most combinations of the dust lane geometry parameters the
observed change is $\ltsim 25$ percent.

\subsubsection{Viewing parameters}

Inclinations of $\sim 45\degrees$ and projected angular separations
between the bar major axis and the disc major axis of $\sim
45\degrees$, such as we used thus far, are favorable orientations for
TW measurements.  In the central panel of Figure~\ref{f:parspace} we
examine the variation in $\omtw / \omint$ as a function of inclination
and bar position angle.  (The dust lane parameters used in this
calculation are \DLxmin\ $=-0.1$, \DLwidth\ $=0.1$ relative to the bar
semi major axis and \DLslope\ $= -0.3$, which correspond to the
approximate values observed in the SBc galaxy NGC~4123 scaled to our
model.)
The error in $\omtw$ reaches a maximum when the bar is close to the
minor axis at extremely low or high inclinations.  Galaxies with such
orientations are not usually selected for TW measurements: at large
inclinations the projected disc on the sky is too narrow to employ the
multiple parallel long-slits required by the TW method while at low
inclinations the observed velocities, $V_{\rm int} \sin i$, will
generally be too small to measure reliably.  Around $i=45\degrees$ and
PA$_{\rm bar} = 45\degrees$, the results are independent of bar
position angle and depend only weakly on inclination.  For $30
\degrees < i < 60\degrees$ and $20\degrees < $ PA$_{\rm bar} <
80\degrees$, these errors are less than $25\%$.

\subsubsection{Scale length parameters}

The observed pattern speed depends only weakly on variations in
$h_{R,{\rm lane}}$. As our initial $h_{R,{\rm lane}}$ is ten times
that of the particle distribution, even large variations around this
initial value do not lead to significant changes in $\omtw$.  The
dependence on $h_{z,{\rm lane}}$ is somewhat stronger as illustrated
in the right panel of figure~\ref{f:parspace}.  Once $h_{R,{\rm
lane}}$ or $h_{z,{\rm lane}}$ become larger than the corresponding
quantities for the stellar distribution, $\omtw / \omint$ becomes
independent of these parameters.

\begin{figure}
  \includegraphics[width=8cm]{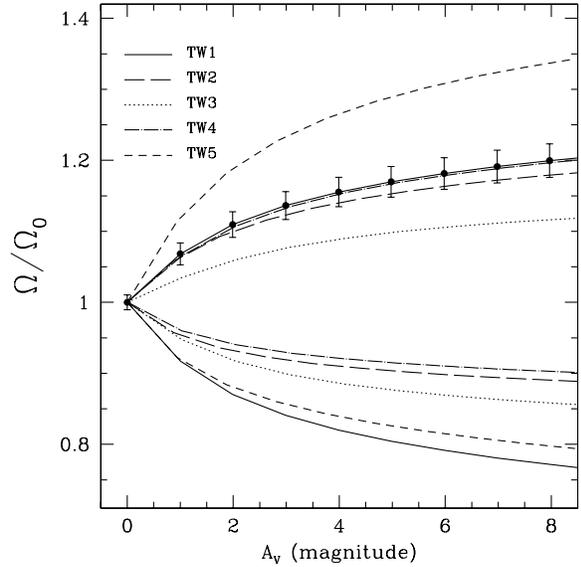}
  \caption{ The ratio $\omtw / \omint$ of the bar pattern speed
  observed with the TW method to the intrinsic bar pattern speed as a
  function of dust lane extinction $A_V$ (defined face-on as described
  in \S \ref{ssec:weights}).  The different curves show the behaviour
  of this ratio for the different $N$-body models.  Two curves are
  shown for each model.  Curves with a ratio $> 1$ are at PA$_{\rm
  bar}$ = $+45\degrees$, while curves with a ratio $< 1$ are at
  PA$_{\rm bar}$ = $-45\degrees$.  The dust lane geometry and dust
  distribution are the same for every model: \DLwidth/$a_B$ = $0.1$,
  \DLslope\ = $-0.3$, \DLxmin/$a_B$ = $-0.1$, $h_{R,{\rm lane}} = 10$
  and $h_{z,{\rm lane}} = 0.1$.  For clarity, the $1\sigma$ errors are
  shown for model TW1 only.
  \label{f:omega}}
\end{figure}

\subsection{Different $N$-body models}

In the previous sections we explored the behavior of $\omtw$ on the
dust model parameters with only model TW1, largely to facilitate
comparisons with D03, who used the same model.  We carried out a
similar parameter space study for all the other $N$-body models
presented in Section~\ref{s:nbody}.  The same dust lane geometry (with
\DLxmin\ and \DLwidth\ rescaled to $\len$) and dust distribution used
in Section~\ref{s:dustlane} was used on each $N$-body model.  Only the
maximum radii of the dust lanes (\DLxmax) differ to account for the
different bar lengths in each model (see Table~\ref{t:snapshots}).
The dust lane widths used here are fairly narrow, about five percent
of the bar semi-major axis length, to resemble the shapes of observed
dust lanes.  These yield qualitatively similar results.  A
quantitative comparison of the different $N$-body models is shown in
Figure~\ref{f:omega} where the ratio $\omtw / \omint$ as a function of
extinction is plotted for each model.  The main difference is in the
errors in model TW5 at positive $PA_{\rm bar}$: the error for this
slow bar is a factor two larger than in the other models.  However at
negative $PA_{\rm bar}$ the error in this model is not much different
than the other models.

\subsection{Star formation}
\label{s:sfres}

\begin{figure}
  \includegraphics[width=8cm]{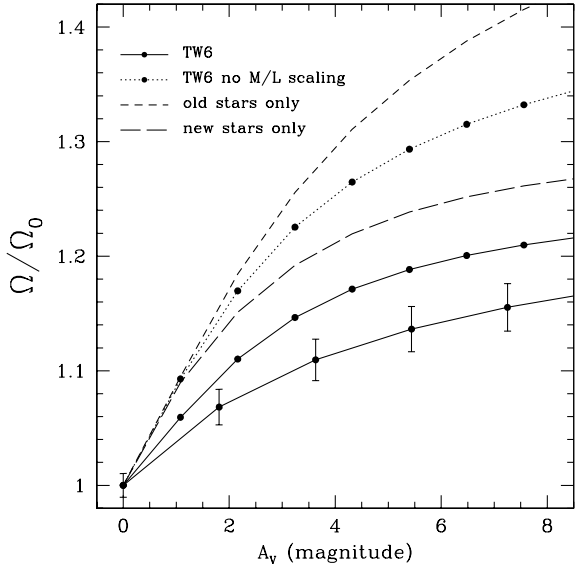}
  \caption{ The ratio $\omtw / \omint$ as a function of dust
  extinction for the model that includes star formation (TW6).  The
  result is not very different from the models shown in
  Fig~\ref{f:omega}. For comparison model TW1 is overplotted as the
  solid line with error bars. Without M/L scaling (see text) dust
  affects model TW6 significantly more than our previous models.  The
  other curves show the behaviour of $\omtw$ in TW6 for particle
  subsets in age. The younger the particles the less the effect of
  dust appears to be.  
  \label{f:omega2}}
\end{figure}

The models that we explored so far do not take star formation into
account. However, in late type systems the SF rate can be large enough
that departures from continuity become too large.  Bright young stars
will contribute disproportionately to the TW integrals $\pin$ and
$\kin$ (see section~\ref{s:twmethod}) and hence may significantly
change the measured pattern speed.  To quantify the effect of star
formation on the TW method we use model TW6.  The result is shown in
Fig.~\ref{f:omega2} and is qualitatively similar to the results
derived from model TW1 to TW5.  Quantitatively, the impact of dust in
TW6 lies between the values in models TW1-TW4 and TW5, with an
$\omtw$ value that differs from its true values (at $A_v = 3$) by 15
percent.  Scaling the intrinsic weights of the particles by their M/L
ratio (section~\ref{s:sf}) is a crucial part of model TW6. As a test
we therefore derive the pattern speeds in this model without any such
scaling (dotted line in Fig.~\ref{f:omega2}).  Without this scaling
the effects of dust are comparable to the slow bar model TW5 with
$\omtw$ deviating from the intrinsic value by 21 percent.

To examine this more closely we divided the particles in two groups,
those particles that were present in the simulation from $t=0$ (`old')
and the ones that were created subsequently by star formation (`new').
The pattern speed derived from younger particles appears to be less
affected by dust. In fact, subdivisions into younger ages show even
shallower curves. Interestingly, the dust free values of $\omint$
derived for the different age bins vary by less than eight percent
from the full model.  This may be somewhat fortuitous as there are
fewer particles at younger ages to derive the TW value.  Their spatial
extent is also smaller as these particles are mostly confined to the
bar region and the nucleus.

This behaviour is probably due to the similar location of dust lanes
and SF regions.  If star formation occurs or is enhanced on the
leading edges of a bar, as is frequently observed, e.g. \citet{she02}
, then absorption by dust may be partially mitigated by the enhanced
luminosity of the young particles.  SF is indeed observed along the
bar in this model.

\section{Discussion}
\label{s:discussion}

The experiments presented in this paper show that dust lanes on the
leading edges of a bar tend to increase the TW derived value of the
bar pattern speed, $\omtw$, when PA$_{\rm bar} > 0$ and decrease it
when PA$_{\rm bar} < 0$.  The changes in $\omtw$ are relatively modest
for realistic dust lanes (i.e. $A_V \ltsim 3$), between 8-25 percent
(cf. Figs \ref{f:omega} and \ref{f:omega2}) and increasing to 20-40
percent at unrealistically large $A_V \sim 8$ or if the bar is
slow.  Somewhat surprisingly, the
simulation that includes star formation (TW6) does not change this
conclusion in a significant way.  Dust lanes alone therefore pose no
serious problem when trying to extend the TW method to late-type
barred galaxies.

\begin{figure*}
 \begin{minipage}{140mm}
  \includegraphics[width=7cm]{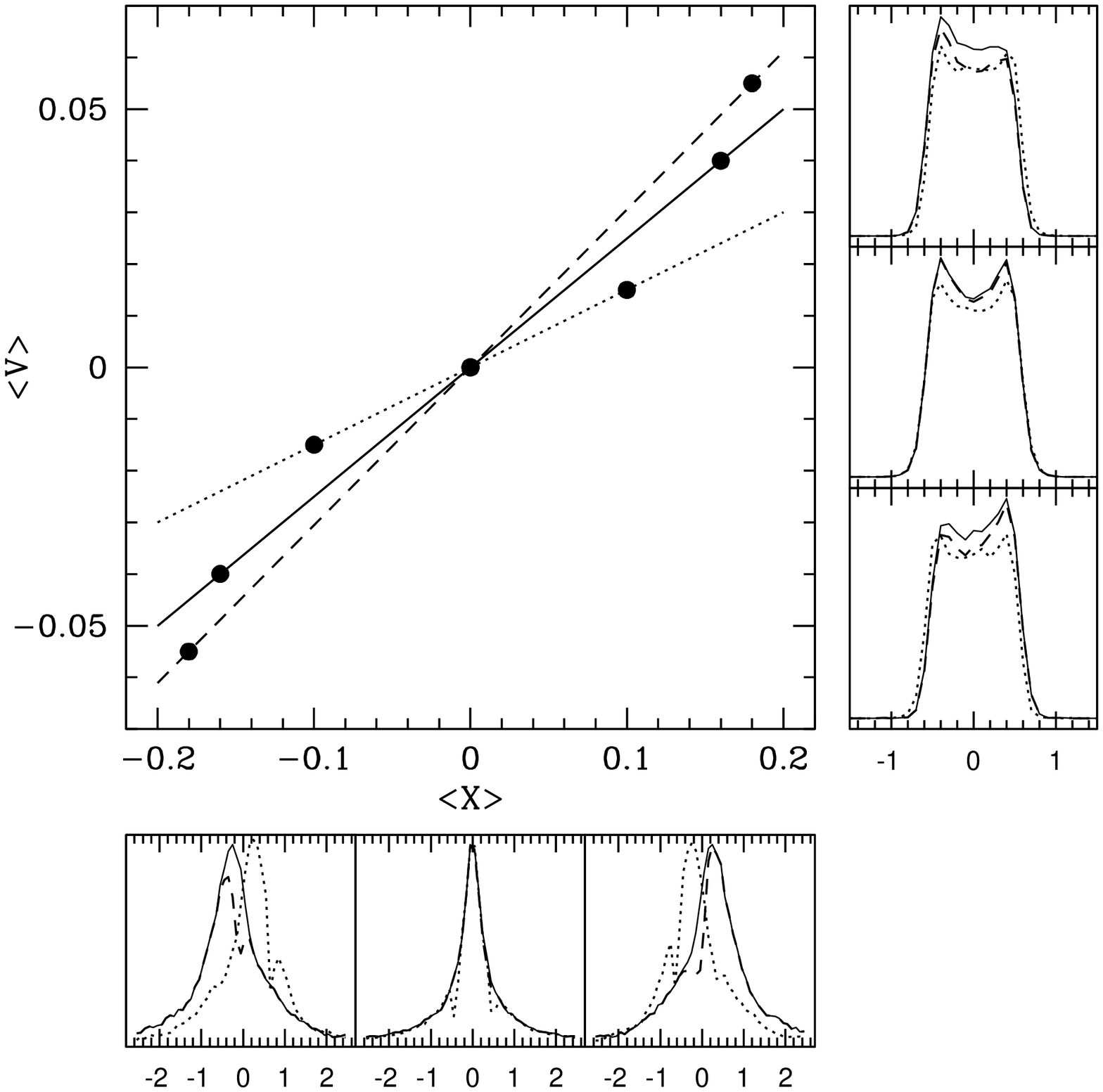}
  \includegraphics[width=7cm]{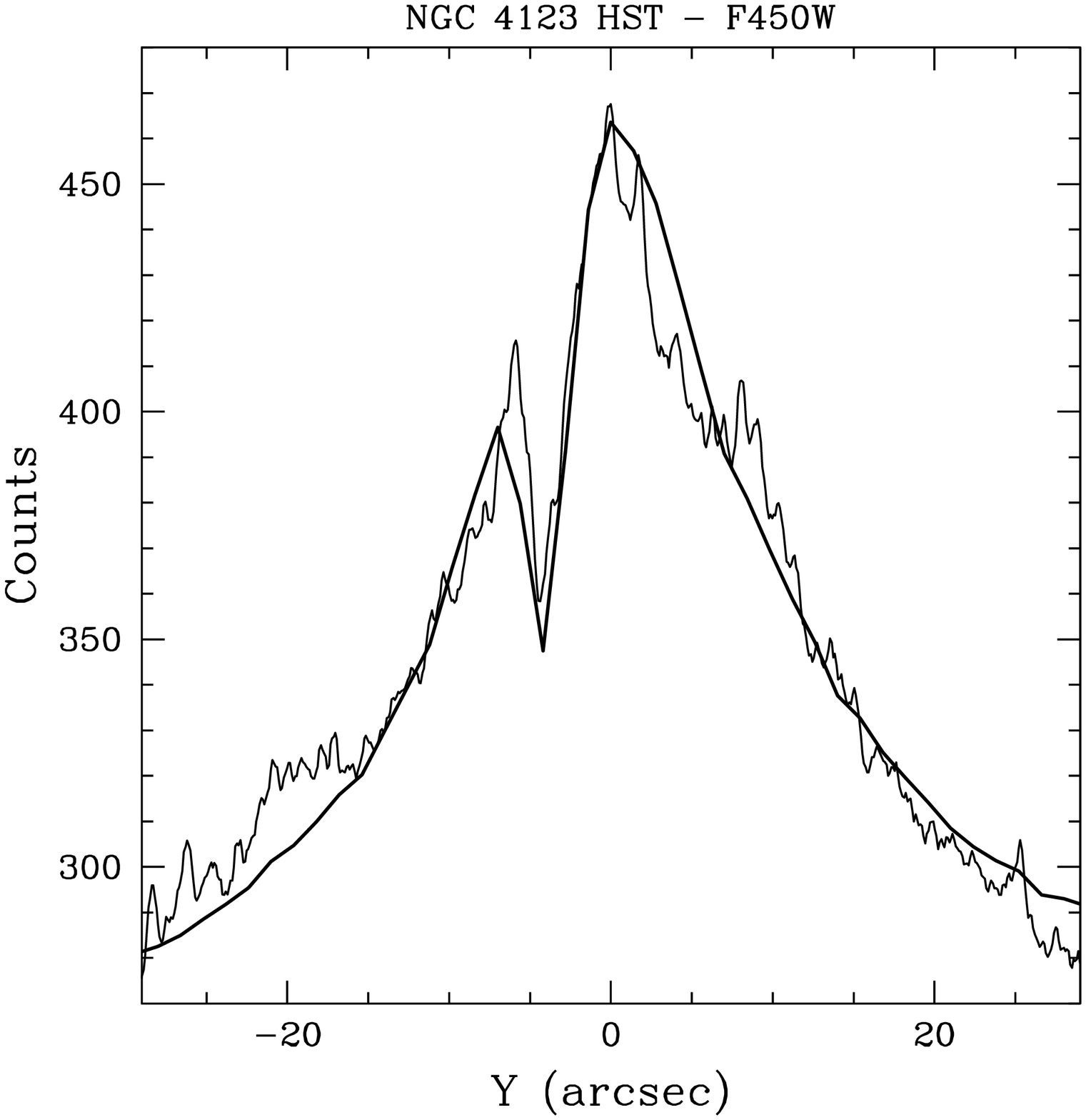}
 \end{minipage}
 \caption{ {\bf left:} Cartoon illustrating the effect of leading edge
  dust lanes on the derived pattern speed. The solid lines show the
  results derived from the $N$-body model using three slits in the
  absence of dust.  The dashed lines show the results obtained when
  leading edge dust lanes are included in a model with PA$_{\rm bar}$
  = $+45\degrees$.  The dotted lines show results derived using the
  same model but with PA$_{\rm bar}$ = $-45\degrees$.  The central
  panel plots the mean line of sight velocities, $\langle V \rangle$
  or $\cal{V}$, versus the mean position, $\langle X \rangle$ or
  $\cal{X}$, for each slit.  The slope of the linear fit gives $\omtw
  \sin i$. The dust lane modifications to the velocity and luminosity
  profiles (shown in the sub panels) lead to either a steeper or a
  shallower slope of the fit depending on PA$_{\rm bar}$ (see
  Figure~\ref{f:dustlane}).  The velocity profiles from top to bottom
  and the luminosity profiles from left to right are derived from
  slits located at $Y=-0.4$, $Y=0.0$, and $Y=0.4$, where $Y$ is
  defined as in Figure \ref{f:dustlane}.
 {\bf right:} The ragged line shows a cut across the dust lane in an
 HST-WFPC2 image of NGC~4123 in the F450W filter.  To improve the
 statistics the profile is summed over 100 pixels ($\sim 10$ arcsec or
 about a fifth of the bar radius) along the dust lane. The dust lane
 itself is clearly noticeable as the depression at $y \approx -5$
 arcsec.  The smooth bold line shows a similar cut in model TW1 at a
 dust lane extinction of $A_V = 3.0$ and summed over approximately the
 same fraction of the dust lane.  The model profile is scaled in an ad
 hoc fashion to match the peak of the observed profile.  The agreement
 with the observed profile is remarkably good.
 \label{f:discussion}}
\end{figure*}

Figure~\ref{f:discussion}a illustrates why the measured $\omtw$
changes when dust lanes are present.  Shown by solid lines in this
figure are the spatial and velocity profiles along three slits in
model TW1 in the absence of dust.  The dashed lines show the same
profiles but for a model that includes leading edge dust lanes when
PA$_{\rm bar} = +45\degrees$, while the dotted lines are for PA$_{\rm
bar} = -45\degrees$.  The spatial profiles clearly show the asymmetric
behaviour expected for this configuration (cf
Figure~\ref{f:dustlane}b).  The measured pattern speed $\omtw$ follows
from the slope of the linear fit to the means of the velocity and
spatial profiles.  The dust distribution affects both the spatial and
the velocity profiles.  If dust would alter only the spatial profiles
but leave the velocity profiles unchanged then the slope of the linear
fit would flatten, leading to slower pattern speed when PA$_{\rm bar}
= +45\degrees$.  However, the associated change in $\kin$ due to the
dust more than offsets this effect and the bar pattern speed therefore
appears larger with dust than without.

The bars in models TW1 to TW4 are classified as `fast'. That is to say
that the ratio of the co-rotation radius to the bar radius in these
models is close to but greater than 1.0. This ratio cannot be smaller
than 1.0 as that would violate self consistency \citep{cp80}.
Applications of the TW method to real data for early-type galaxies
have not revealed slow (i.e. with a ratio $\gtrsim 2$) bars.  We
nevertheless included a slow bar, model TW5, to examine the effect of
dust in such a case.  In this model the impact of dust on $\omtw$ is
larger than in the fast bars (Fig.~\ref{f:omega}) This is not
surprising as the absolute change in $\Omega$ between successive steps
in $A_v$ is, to within a factor two, independent of the details of a
particular model.  In other words, the factor by which dust acts to
change $\omtw$ is similar for all models and hence makes more of an
impact on a slow bar.  A similar behaviour is observed in model TW6
(Fig.~\ref{f:omega2}).  The ratio of bar radius to corotation radius
is about 1.5 indicating that this bar is slow-ish. The observed
pattern speed is indeed more affected by dust than in models TW1-TW4.

An important concern with TW measurements in late-type systems is the
requirement of a well constrained PA.  \citet{deb03} demonstrates that
a PA error of $5 \degrees$ can result in pattern speed errors by up to
100 percent \citep[see also][]{dw04}.  In real galaxies,
non-axisymmetric features can lead to uncertainties in the derived
position angles (and inclinations) that are on average $5 \degrees$
\citep{bs03}. It is therefore crucial to select systems with well
constrained PAs when applying the TW method to real late-type
galaxies.

\citet{wei01} presented a $B-I$ color map of the SBc galaxy NGC~4123
that clearly delineates the dust lanes on the leading edges of its
bar.  They found that $\left<B-I\right> = 2.3$ in the
dust lanes.  With the extinction law of \citet{rl85} this colour
implies an average extinction in the dust lanes of $A_V = 2.7$
magnitude.
We checked this value using an $HST$ archival WFPC2 F450W band image
of NGC~4123 (SNAP-9042 PI: Smartt) to compare a surface brightness
profile across the dust lane (i.e. parallel to the bar minor axis) in
the $HST$ image to a similarly placed slit in model TW1.  We use $A_V
= 3$ to match the value implied by the \citet{wei01} colour map.  As
the exposure time of the $HST$ image is only 160s we increase the
$S/N$ of the surface brightness cut by summing over 100 pixels along
the dust lane.  This corresponds to a length of about 10 arcsec over
the dust lane or a fifth of the bar radius (Weiner \etal\ 2001).  The
ragged line in Figure~\ref{f:discussion}b shows the resulting surface
brightness profile.  The dust lane itself is clearly noticeable as the
depression near $y \approx -5$ arcsec.  The smooth bold line
represents a similar cut along the model image averaged over a similar
fraction of the dust lane.  The model profile is scaled ad hoc to
match the peak in the NGC~4123 data: the resulting profile fits the
observed one remarkably well.
Using this extinction, the models shown in Figures~\ref{f:omega} \& 6
predict a $\sim 15$ percent change in the observed bar pattern speed.
If, on the other hand the observations were carried out in $K$-band
using the CO bandhead, the extinction of $A_V = 2.7$ magnitudes
translates to $A_K = 0.112$ \citep[e.g.][their table 3]{rl85}.  This
extinction would lead to insignificant errors in $\omtw$ from
obscuration.

Our results suggest that the application of the TW method can be
extended to later-type barred galaxies.  We therefore propose that at
this time a test case of a TW measurement in a late-type galaxy is
warranted.  A useful approach would be to try a TW measurement on a
bar with pattern speed measured by hydrodynamical modeling
\citep{wei01, per04}.

\section*{Acknowledgments}
V.P.D. is supported by a Brooks Prize Fellowship at the University of
Washington and receives partial support from NSF ITR grant
PHY-0205413.  We would like to thank Rok Ro\v skar for granting us
permission to use model TW5 here ahead of publication.  We thank the
anonymous referee for comments that helped to improve this paper.

\section*{Appendix}

We know the mass $M_p$ and time of formation $t_p$ of all particles in
model TW6.  To calculate the luminosity of a particle we follow
Tinsely (1973). We begin by assuming that the stars that make up a
particle (typical mass $\sim 10^5 M_\odot$) are distributed according
to a Salpeter IMF: $f(M) = C M^{-2.35}$.  The number of stars in a
particle, $C$, can be found by setting $M_{\rm tot} \times $C$ = M_p$.
The total mass is
\[
M_{\rm tot} = \int_{M_{lo}}^{M_{up}} f(M) dM 
\]
where $M_{lo} = 0.1 \ M_\odot$ and $M_{up} = 100 \ M_\odot$.  We can
now calculate the total luminosity of this particle by assuming that
the luminosity along the main sequence scales as $M^\alpha$ where
$\alpha = -4.1$. Then,
\[
L_{\rm tot} = C \int_{M_{lo}}^{M_{up}}{M^{\alpha} M^{-2.35} \ dM}
\]
at the moment the particle is created.

At later times the more massive stars will have turned off the main
sequence. The upper limit for a star on the main sequence is given by
$M_{up} = \frac{t_p}{{\rm 10 Gyr}}^{1/\gamma}$ where $\gamma = -2.5$.
By using this value for $M_{up}$ in the $L_{\rm tot}$ equation the
luminosity of a particle at any time $t$ can be derived.  Stars with
masses larger than $M_{up}$ go through a luminous but short lived
giant phase.  We include the contribution of giants in our calculation
of $L_{\rm tot}$ following \citet{kru89}. But only in the older
particles do they make a contribution comparable to the main sequence
stars.

\label{lastpage}

\end{document}